\documentclass[fleqn,usenatbib]{mnras}

\usepackage{newtxtext,newtxmath}
\usepackage{hyperref}
\usepackage[T1]{fontenc}

\DeclareRobustCommand{\VAN}[3]{#2}
\let\VANthebibliography\thebibliography
\def\thebibliography{\DeclareRobustCommand{\VAN}[3]{##3}\VANthebibliography}

\usepackage{graphicx}	% Including figure files
\usepackage{amsmath}	% Advanced maths commands
\usepackage[nolist]{acronym}
\usepackage{tablefootnote}
\usepackage{array}
\usepackage{multirow,multicol}
\usepackage{comment}
\usepackage{xspace}

\newcommand{\ergcm}[1]{$\times 10^{#1}$ erg cm$^{-2}$ s$^{-1}$\xspace}

\newcommand{\ergs}[1]{$\times 10^{#1}$ erg s$^{-1}$\xspace}
\newcommand{\oergs}[1]{$10^{#1}$ erg s$^{-1}$\xspace}

\newcommand{\ohcm}[1]{$10^{#1}$ cm$^{-2}$\xspace}

\newcommand{\expo}[1]{$\times 10^{#1}$\xspace}

\newcommand{\nhsmc}{N$_{\rm H}^{\rm SMC}$\xspace}

\newcommand{\nhGal}{N$_{\rm H}^{\rm Gal}$\xspace}

\newcommand{\ltsima}{$\buildrel < \over \sim$}
\newcommand{\lsim}{\lower.5ex\hbox{\ltsima}}
\newcommand{\gtsima}{$\buildrel > \over \sim$}
\newcommand{\gsim}{\lower.5ex\hbox{\gtsima}}

\newcommand{\xspec}{{\tt XSPEC}\xspace}

\newcommand{\pattern}{{\tt PATTERN}\xspace}

\newcommand{\swift}{{\it Swift}\xspace}
\newcommand{\xmm}{{\it XMM-Newton}\xspace}
\newcommand{\EP}{{\it EP}\xspace}
\newcommand{\EPFXT}{{{\it EP}-FXT}\xspace}
\newcommand{\fullEP}{{\it Einstein Probe}\xspace}

\newcommand{\ROSAT}{{\it ROSAT}\xspace}

\newcommand{\ero}{\mbox{eROSITA}\xspace}
\newcommand{\src}{\mbox{EP\,J005146.9$-$730930}\xspace}
\newcommand{\sxpsrc}{\mbox{SXP\,146.8}\xspace}
\newcommand{\oglesrc}{\mbox{OGLE\,J005147.58$-$730924.7}\xspace}
\newcommand{\nearsxp}{\mbox{SXP\,172}\xspace}

\title[EP\,J005146.9$-$730930, a new Be/X-ray binary]{
Discovery of 146.8\,s pulsations from EP\,J005146.9$-$730930, a new transient Be/X-ray binary in the SMC}

\author[F. Haberl et al.]{F. Haberl$^{1}$\thanks{E-mail: fwh@mpe.mpg.de},
Y. Xu$^{2}$\thanks{E-mail: xuyj@ihep.ac.cn},
C. Maitra$^{1}$,
G. Vasilopoulos$^{3,4}$,
W. Zhang$^{5}$,
W. Yuan$^{5,6}$,
C. Jin$^{5,6,7}$,
H.N. Yang$^{5,1}$,
\newauthor
L. Ducci$^{8}$,
D.M. Kaltenbrunner$^{1}$,
P. Maggi$^{9}$,
A. Rau$^{1}$
\\
$^{1}$Max-Planck-Institut f{\"u}r extraterrestrische Physik, Gie{\ss}enbachstra{\ss}e 1, 85748 Garching, Germany\\
$^{2}$Key Laboratory of Particle Astrophysics, Institute of High Energy Physics, Chinese Academy of Sciences, Beijing 100049, China\label{ihep}\\
$^{3}$Department of Physics, National and Kapodistrian University of Athens, University Campus Zografos, GR 15784, Athens, Greece\label{nku}\\
$^{4}$Institute of Accelerating Systems \& Applications, University Campus Zografos, Athens, Greece\label{iasa}\\
$^{5}$National Astronomical Observatories, Chinese Academy of Sciences, Beijing, 100101, People's Republic of China\label{cas}\\
$^{6}$School of Astronomy and Space Sciences, University of Chinese Academy of Sciences, Beijing, 100049, People's Republic of China\label{ucas}\\
$^{7}$Institute for Frontier in Astronomy and Astrophysics, Beijing Normal University, Beijing, 102206, People's Republic of China\label{bnu}\\
$^{8}$Institut f{\"u}r Astronomie und Astrophysik, Sand 1, 72076 T{\"u}bingen, Germany\label{iaat}\\
$^{9}$Universit\'e de Strasbourg, CNRS, Observatoire astronomique de Strasbourg, UMR 7550, 67000 Strasbourg, France\label{oas}\\
} 

\date{Accepted 2025 July 22. Received 2025 July 22; in original form 2025 April 20}

\pubyear{2025}
\begin{document}
\label{firstpage}
\pagerange{\pageref{firstpage}--\pageref{lastpage}}
\maketitle


\begin{abstract}
   Recent observations of the Small Magellanic Cloud (SMC) with \fullEP (\EP) revealed a new transient X-ray source, most likely identified as Be/X-ray binary.
   To characterise the X-ray properties of \src and in particular to look for pulsations in the X-ray flux, we triggered an \xmm anticipated target of opportunity observation. To follow the flux evolution during the outburst we monitored the source for about three months with the Follow-up X-ray Telescope of \EP.
   The \xmm observation was performed on 2024 September 15 and we used the data from the European Photon Imaging Camera (EPIC) for detailed spectral and timing analyses.
   The EPIC X-ray spectrum is well described by an absorbed power law with photon index of 1.25 $\pm$ 0.04 and the timing analysis revealed pulsations with 146.79 $\pm$ 0.03\,s. The source flux had decreased by a factor of about 10 since the observed maximum about one month before the \xmm observation. \src was never detected significantly during serendipitous observations before September 2024. The characteristics of the X-ray brightening suggest the source was discovered during a type II outburst reaching an X-ray peak luminosity of $\sim$$2\times10^{37}$\,erg s$^{-1}$. The trend of spectral hardening towards higher luminosities observed by \EP suggests that the source is accreting below the critical luminosity, yielding an estimated lower limit for the pulsar magnetic field strength of $3.3\times10^{12}$\,G. The improved X-ray position confirms the candidate Be star \oglesrc as optical counterpart.
   We conclude that \src = \sxpsrc is a new Be/X-ray binary pulsar in the SMC.
\end{abstract}

\begin{keywords}
galaxies: individual (SMC) -- 
X-rays: binaries --
stars: emission-line, Be -- 
stars: neutron
pulsars: individual (\src)
\end{keywords}

%%%%%%%%%%%%%%%%%%%%%%%%%%%%%%%%%%%%%%%%%%%%%%%%%%

%%%%%%%%%%%%%%%%% BODY OF PAPER %%%%%%%%%%%%%%%%%%ß

\section{Introduction}
\label{sec:intro}

The Small Magellanic Cloud (SMC) is well known for the large population of high-mass X-ray binaries (HMXBs) with more than 125 systems known. About half of them revealed X-ray pulsations indicating the spin period of an accretion-powered neutron star. 
Since the compilation of the SMC HMXB catalogue by \citet{2016A&A...586A..81H}, seven new pulse periods from known or new Be/X-ray binaries (BeXRBs, the major subclass of HMXBs) were discovered \citep{2016ATel.9229....1V,2019ApJ...884....2L,2020MNRAS.494.5350V,2019ATel13312....1H,2020ATel13823....1K,2023ATel16321....1C,2023ATel15886....1M}, increasing the number of BeXRB pulsars to 70.
The large number of HMXBs is most likely related to the recent star formation history of the SMC which shows a maximum around 40\,Myr ago. At this age the maximum occurrence of the Be phenomenon is expected \citep{2010ApJ...716L.140A}.

BeXRBs are often discovered during X-ray outbursts. Periodic type-I outbursts (peak X-ray luminosities $<$\oergs{37}) occur around the periastron passage of the neutron star and tend to last only a small fraction of the binary period. Type-II outbursts are usually stronger ($>$\oergs{37}) and can last longer than the orbital period. For more details see the review by e.g. \citet{2011Ap&SS.332....1R}.

During recent monitoring observations of the SMC, a new transient source was discovered in data of the Follow-up X-ray Telescope (FXT) on board the newly launched \fullEP mission \citep[\EP;][]{2022hxga.book...86Y}. The source, \src, was first detected by \EPFXT on 2024 July 25 at an 0.5$-$10\,keV X-ray flux of 6.1\ergcm{-13}. \src significantly brightened until August 2 (1.1\ergcm{-11}) and August 15 (3.5\ergcm{-11}). It remained near this flux level until August 29 (1.8\ergcm{-11}) as reported by \citet{2024ATel16795....1X}.
The flux increase was also seen during monitoring observations with the Neil Gehrels Swift observatory \citep[\swift;][]{2004ApJ...611.1005G}. \src was detected in the data of the X-ray telescope (XRT) with a maximum count rate on August 13. From the improved X-ray position the candidate Be star \oglesrc was proposed as optical counterpart \citep{2024ATel16796....1K}.

In order to look for pulsations we triggered one of our pre-approved anticipated target of opportunity observations with \xmm. In Section\,\ref{sec:xray} we describe the observations and the data analysis. The results are discussed in Section\,\ref{sec:discussion}.

\section{X-ray observations and data analysis}
\label{sec:xray}

\subsection{\fullEP}

The outburst of \src was found by \EP during a regular monitoring program of the SMC using FXT. \EP observed \src with multiple snapshots from 2024 July 25th to October 22nd (see Table \ref{tab:fxt} for details), then further observations of the corresponding SMC regions were prohibited due to solar constraints.  \EPFXT consists of two identical and co-aligned telescope units, FXTA and FXTB, both observed \src in the full-frame (FF) mode with thin filters.

We reduced the \EPFXT data following standard procedures using the FXT Data Analysis Software (\texttt{FXTDAS}) v1.10 with CALDB v1.10. The source spectra were extracted using a circular region with a radius of 50\arcsec\ centered on the source position from FXTA and FXTB, respectively. Corresponding background spectra were selected from source-free areas with a radius of 300\arcsec\ within the field-of-view. Due to the short exposure times of the individual \EPFXT monitoring observations, no periodicity was significantly detected in any of the single-epoch observations. Therefore, we only report the energy and flux evolution measured by \EP in this paper. \src and the nearby BeXRB \nearsxp (see below) could be resolved by FXT (on-axis half-power diameter spatial resolution of $\lesssim$30\arcsec) but not by the Wide-field X-ray Telescope (WXT) onboard \EP. Therefore, in this paper we only report the X-ray fluxes measured by FXT. For spectral analysis, we group the energy spectra to have a minimum of one count in each bin, and jointly fit the FXTA and FXTB spectra in \xspec \citep{1996ASPC..101...17A} using C-statistics \citep{cash79} to determine the best-fit parameters, which is preferable to $\chi^2$-statistics for low-count spectra.

%-------------------------
\begin{table*}
\centering
\caption[]{\EPFXT monitoring observations of \src with the best-fit spectral parameters. Both FXTA and FXTB data were taken in the FF mode, pile-up effects were negligible at such flux levels \citep{zhang25}. The best-fit spectral parameters of N$_{\rm H}^{\rm SMC}$ and $\Gamma$ are obtained from joint fits of FXTA and FXTB. The net count rates and the observed X-ray flux in the 0.2--10 keV band listed are the average values obtained by FXTA and FXTB. All uncertainties quoted are at the 90\% confidence level.}
\label{tab:fxt}
\begin{tabular}{c|cccc|ccccc}
\hline\hline\noalign{\smallskip}
\multicolumn{1}{l|}{No.} &
\multicolumn{1}{c}{ObsID} &
\multicolumn{1}{c}{Observation time} &
\multicolumn{1}{c}{Exposure} &
\multicolumn{1}{c|}{Net count rate} &
\multicolumn{1}{c}{N$_{\rm H}^{\rm SMC}$} &
\multicolumn{1}{c}{$\Gamma$} &
\multicolumn{1}{c}{Flux$_{\rm 0.2-10\,keV}$} &
\multicolumn{1}{c}{Cstat/dof} \\
&  & 
\multicolumn{1}{c}{(UT)} &
\multicolumn{1}{c}{(s)}  &
\multicolumn{1}{c|}{($10^{-2}$ s$^{-1}$)} & 
\multicolumn{1}{c}{(10$^{21}$ cm$^{-2}$)}  &    &
\multicolumn{1}{c}{(erg~cm$^{-2}$~s$^{-1}$)}   & \\
\noalign{\smallskip}\hline\noalign{\smallskip}
1  &11908398338    &2024-07-25 15:59 -- 07-25 16:18     &1148     &${3.0\pm0.4}$     &<9.7                    &$0.7_{-0.3}^{+0.6}$   &${(2.3\pm{0.6})}{\times}10^{-12}$     &70.9/61        \\ 

2  &11908398339    &2024-08-01 23:05 -- 08-01 23:23     &1075     &${20.6\pm1.0}$        &$13\pm{5}$              &$1.1\pm{0.2}$   &${(1.21\pm0.11)}{\times}10^{-11}$    &271.0/328      \\
 
3  &11908428547    &2024-08-02 23:08 -- 08-02 23:29      &1288     &${23.1\pm0.9}$        &$9.8_{-3.8}^{+4.3}$     &$1.0\pm{0.2}$   &${(1.28\pm0.10)}{\times}10^{-11}$            &312.0/412      \\ 

4  &11908398374    &2024-08-14 17:24 -- 08-14 17:56     &979      &${55.5\pm1.7}$        &$2.3_{-1.5}^{+1.7}$     &$0.7\pm{0.1}$   &${(3.84\pm0.21)}{\times}10^{-11}$          &486.5/595      \\
 
5  &11908428567    &2024-08-15 22:56 -- 08-15 23:19     &1405     &${63.5\pm1.5}$        &$3.1_{-1.3}^{+1.5}$     &$0.8\pm{0.1}$   &${(3.66\pm0.15)}{\times}10^{-11}$        &699.4/767      \\ 
 
6  &11908525313    &2024-08-21 18:27 -- 08-21 18:53    &1545     &${39.9\pm1.1}$        &$3.9_{-1.7}^{+1.9}$     &$0.9\pm{0.1}$   &${(2.53\pm0.13)}{\times}10^{-11}$          &565.2/617      \\ 

7  &11908553988    &2024-08-28 16:41 -- 08-28 16:57     &969      &${25.0\pm1.1}$        &$6.7_{-3.2}^{+3.7}$     &$1.0\pm{0.2}$   &${(1.42\pm0.12)}{\times}10^{-11}$         &284.5/340       \\ 
 
8  &11908561156    &2024-08-29 18:49 -- 08-29 19:12     &1306     &${33.9\pm1.1}$        &$3.5_{-2.1}^{+2.5}$     &$0.8\pm{0.1}$   &${(1.98\pm0.12)}{\times}10^{-11}$           &456.1/529       \\ 
 
9  &11908572422    &2024-09-03 17:19 -- 09-03 17:47      &1648     &${21.3\pm0.8}$        &$4.9_{-2.7}^{+3.0}$     &$0.9\pm{0.2}$   &${(1.37\pm0.10)}{\times}10^{-11}$          &355.2/451       \\ 
 
10 &11908609289    &2024-09-12 17:23 -- 09-12 17:44    &1275     &${10.6\pm0.6}$        &$15_{-7}^{+9}$          &$1.6_{-0.3}^{+0.4}$   &${(3.6\pm0.5)}{\times}10^{-12}$     &164.3/222       \\
 
11 &11908656897    &2024-09-26 19:49 -- 09-26 20:09   &1224     &${6.8\pm0.5}$     &$15_{-8}^{+10}$         &$1.6\pm{0.5}$   &${(2.3\pm0.4)}{\times}10^{-12}$     &136.9/143       \\ 

12 &11908663553    &2024-09-28 22:45 -- 09-28 23:27     &1571     &${6.3\pm0.5}$     &<9.3                   &$1.0_{-0.3}^{+0.4}$   &${(3.5\pm0.5)}{\times}10^{-12}$     &137.8/170        \\

13 &11908712722    &2024-09-26 19:49 -- 09-26 20:09   &867      &${4.4\pm0.5}$     &<8.5                    &$0.7_{-0.3}^{+0.6}$   &${(3.5\pm0.8)}{\times}10^{-12}$     &57.6/68          \\                                                                                   
14 &11908731657    &2024-10-16 18:05 -- 10-16 18:30   &1459     &${2.7\pm0.3}$     &<15.1                    &$1.0_{-0.5}^{+0.6}$   &${(1.4\pm0.3)}{\times}10^{-12}$    &77.9/73          \\                                                                                  
15 &11908753155    &2024-10-10 23:55 -- 10-11 00:09   &1104     &${1.7\pm0.3}$     &<11.5                    &$0.6_{-0.5}^{+0.7}$   &${(1.4\pm0.6)}{\times}10^{-12}$      &40.3/32          \\                           

16 &11908754179    &2024-10-22 08:23 -- 10-22 08:39   &971      &${2.3\pm0.4}$     &<13.9                    &$1.7_{-0.5}^{+1.2}$   &${(5.7\pm1.9)}{\times}10^{-13}$    &23.2/38          \\
\noalign{\smallskip}\hline
\end{tabular}
\end{table*}
%-------------------------

%%%%% EP spectral analysis
\subsubsection{Spectral analysis}
\label{sec:FXTspec}

\begin{figure}
\centering
 \resizebox{\hsize}{!}{\includegraphics{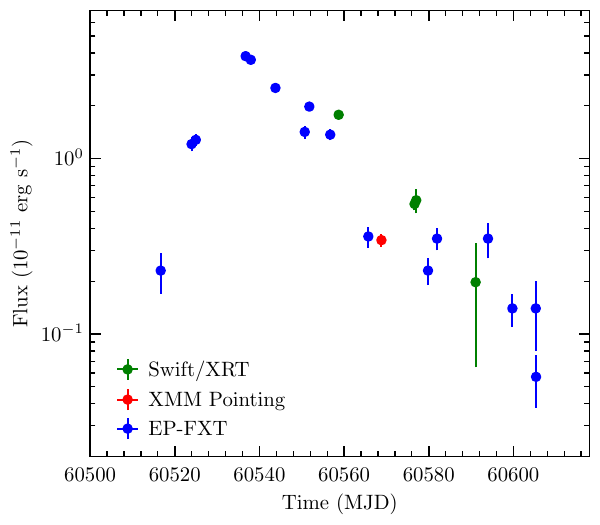}}
  \caption{X-ray flux (observed, 0.2$-$10.0 keV) of \src during the \EPFXT monitoring of the outburst. The flux derived from the \xmm (Section\,\ref{sec:EPICspec}) and \swift (Section\,\ref{sec:ltflux}) observations are included.}
  \label{fig:epflux}
\end{figure}

\begin{figure*}
\centering
 \resizebox{\hsize}{!}{\includegraphics{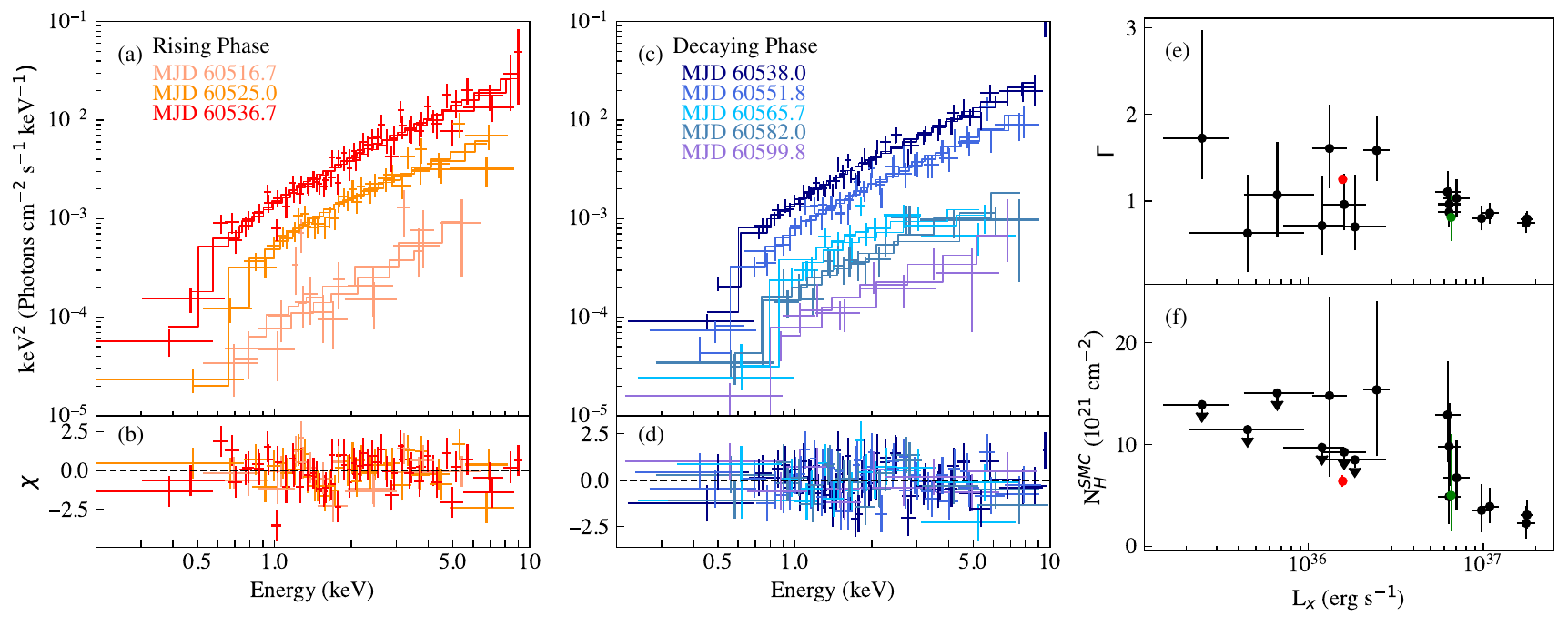}}
  \caption{Left: \EPFXT energy spectra folded with the best-fit models during the rising phase of the outburst (a), along with the spectral residuals (b). FXTA and FXTB spectra taken in a single epoch are plotted in the same color, and are rebinned for visual clearity. Middle: Sample \EPFXT spectra during the decaying phase of the outburst. Right: Relation of the photon index (e) and the intrinsic absorption column density in SMC along the line-of-sight of \src (f) with the absorption-corrected X-ray luminosity in 0.2--10 keV, assuming a source distance of 62 kpc. Measurements by \EP, \xmm, and \swift are plotted in black, red, and green, respectively.}
  \label{fig:epspec}
\end{figure*}

To study the spectral evolution of \src during the outburst, we performed a spectral modeling of the multi-epoch observations taken by \EPFXT. We jointly fit the FXTA and FXTB spectra in the 0.2--10 keV band with an absorbed power-law model plus a cross-normalization constant: \texttt{constant*tbabs*tbvarabs*powerlaw}. The cross-normalization constant is allowed to vary freely for FXTB and is assumed to be unity for FXTA. 
For the absorption along the line of sight we used two components, 
one accounting for the foreground absorption in the Galaxy with abundances in the interstellar medium (ISM) following \citet{2000ApJ...542..914W} and atomic cross-sections from \citet{1996ApJ...465..487V}. 
The Galactic column density, \nhGal, was taken from \citet{1990ARA&A..28..215D} and fixed in the fit at 0.56$\times10^{21}$ cm$^{-2}$.
For the absorption local to the source and in the ISM of the SMC, we assumed elemental abundances of 1.0 for He and 0.2 solar for elements with Z > 2 \citep{1992ApJ...384..508R} and left the column density free in the fit. 
The best-fit parameters measured during the \EPFXT monitoring observations, including the intrinsic absorption column density in the SMC, \nhsmc, the power-law photon index, $\Gamma$, the X-ray flux in the 0.2--10 keV band, and the fitting statistics are listed in Table \ref{tab:fxt}. 
As seen from the C-statistics, the model describes the FXT energy spectra well. The flux evolution during the X-ray outburst is plotted in Figure\,\ref{fig:epflux}.
The \EPFXT spectra together with the evolution of \nhsmc and $\Gamma$ as function of X-ray luminosity are shown in Figure\,\ref{fig:epspec}.
The \EPFXT observation on MJD 60565.7 was conducted about three days before the \xmm observation. The best-fit parameters determined by \EPFXT on MJD 60565.7 could be considered as roughly consistent with the \xmm results (see below) within errors, given that the \EPFXT parameters are less well constrained due to poorer statistics. 

%--------------------------
%--------------------------

%--------------------------
%--------------------------

\subsection{\xmm}

%--------------------------
\begin{table*}
\centering
\caption[]{The \xmm observation of \src. The net exposure times used for the EPIC spectra and images after background-flare screening are listed for pn, MOS1 and MOS2, respectively. }
\label{tab:xmmobs}
\begin{tabular}{lrlcc}
\hline\hline\noalign{\smallskip}
\multicolumn{1}{l}{Instrument} &
\multicolumn{1}{c}{Observation} &
\multicolumn{1}{c}{Observation} &
\multicolumn{2}{c}{Net exposure time} \\
\multicolumn{1}{l}{} &
\multicolumn{1}{c}{ID} &
\multicolumn{1}{c}{time (UT)} &
\multicolumn{1}{c}{(ks)} &
\multicolumn{1}{c}{Fraction of total} \\
\noalign{\smallskip}\hline\noalign{\smallskip}
EPIC  & 0903450101  & 2024-09-15 16:19 -- 22:46  & 9.00, 13.25, 13.22 & 67\%, 77\%, 76\% \\
\hline
\end{tabular}
\end{table*}
%--------------------------
%--------------------------
\begin{table*}
\centering
\caption[]{Astrometric alignments.}
\label{tab:xmmastrometry}
\begin{tabular}{lcccccccc}
\hline\hline\noalign{\smallskip}
\multicolumn{1}{l}{X-ray source} &
\multicolumn{2}{c}{R.A. (J2000) Dec.} &
\multicolumn{1}{c}{1$\sigma$\,Err} &
\multicolumn{2}{c}{R.A. (J2000) Dec.} &
\multicolumn{2}{c}{R.A. (J2000) Dec.} &
\multicolumn{1}{c}{1$\sigma$\,Err} \\
\multicolumn{1}{l}{} &
\multicolumn{2}{c}{EPIC uncorrected} &
\multicolumn{1}{c}{} &
\multicolumn{2}{c}{Gaia DR3} &
\multicolumn{2}{c}{EPIC corrected} &
\multicolumn{1}{c}{} \\
\multicolumn{1}{l}{} &
\multicolumn{1}{c}{(h m s)} &
\multicolumn{1}{c}{(\degr\ \arcmin\ \arcsec)} &
\multicolumn{1}{c}{(\arcsec)} &
\multicolumn{1}{c}{(h m s)} &
\multicolumn{1}{c}{(\degr\ \arcmin\ \arcsec)} &
\multicolumn{1}{c}{(h m s)} &
\multicolumn{1}{c}{(\degr\ \arcmin\ \arcsec)} &
\multicolumn{1}{c}{(\arcsec)} \\
\noalign{\smallskip}\hline\noalign{\smallskip}
\src     & 00 51 47.695 & -73 09 25.60 & 0.51 & 00 51 47.584 & -73 09 25.15 & 00 51 47.59 & -73 09 25.1 & 0.1 \\
\nearsxp & 00 51 52.126 & -73 10 34.66 & 0.50 & 00 51 52.025 & -73 10 34.14 & 00 51 52.03 & -73 10 34.1 & 0.1 \\
\hline
\end{tabular}
\end{table*}
%--------------------------

\xmm carries three X-ray telescopes, two equipped with Metal Oxide Semi-conductor (MOS) CCD arrays \citep{2001A&A...365L..27T} and one with a pn-CCD \citep[][]{2001A&A...365L..18S}. These detectors, which comprise the  European Photon Imaging Camera (EPIC), are sensitive in the 0.15$-$12\,keV energy band. 
The \xmm observation was performed on 2024 September 15 (for details see Table\,\ref{tab:xmmobs}).

\begin{figure}
\centering
 \resizebox{\hsize}{!}{\includegraphics{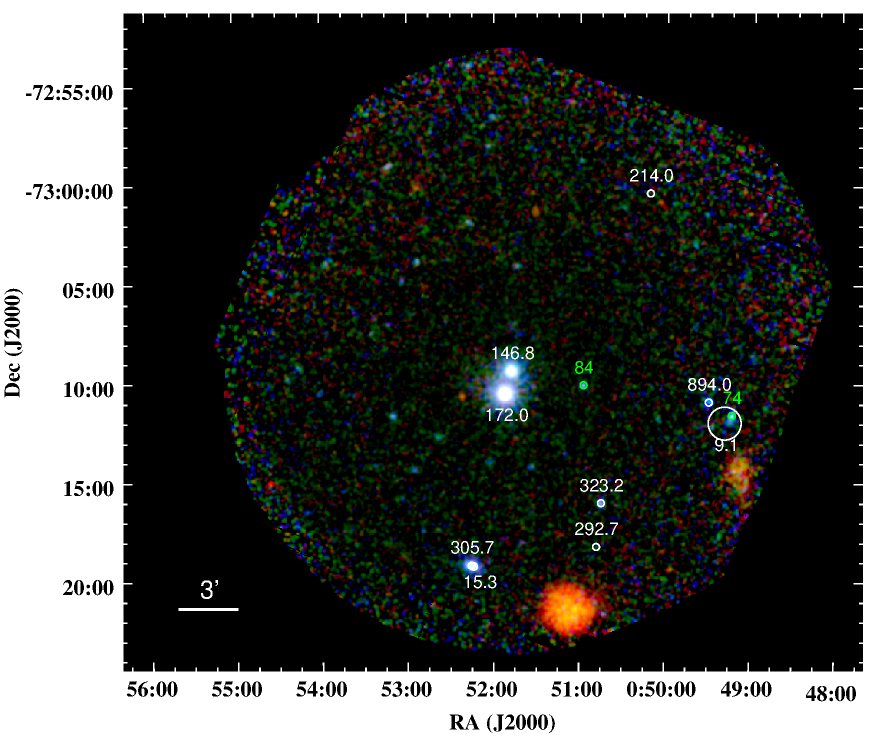}}
  \caption{\xmm EPIC colour image. Red, green and blue represent the detector background-subtracted and vignetting-corrected count rate images in the bands 0.2$-$1.0\,keV, 1.0$-$2.0\,keV and 2.0$-$4.5\,keV, respectively. HMXBs from \citet{2016A&A...586A..81H} known in the field are marked with their pulse period (white), if known, or catalogue numbers otherwise. An association of the 9.1\,s pulsar with source 74 still needs confirmation.}
  \label{fig:EPICima}
\end{figure}

We used the \xmm Science Analysis Software (SAS)\,21.0\footnote{\url{https://www.cosmos.esa.int/web/xmm-newton/sas}}
package to process the EPIC data.
To exclude time intervals of high background during soft-proton flares, we filtered out time intervals with background rates in the 7.0$-$15.0\,keV band above 8 and 2.5\,cts ks$^{-1}$ arcmin$^{-2}$ for EPIC-pn and EPIC-MOS, respectively. 
The X-ray position of \src was determined using the standard \xmm source detection task {\tt edetect\_chain} on the EPIC images after screening for the background flaring. The uncertainty is dominated by the remaining systematic error of 0.5\arcsec. The position is 1.1\arcsec\ from \oglesrc, the optical counterpart proposed by \citet{2024ATel16796....1K} and 0.63\arcsec\ from the Gaia DR3 position \citep{2022yCat.1355....0G} of the counterpart. During the \xmm observation the nearby BeXRB \nearsxp \citep[$\sim$70\arcsec\ angular distance;][]{2010ApJ...716.1217L,2009ApJ...707.1080A} was a factor of $\sim$2.5 brighter than \src (see Figure\,\ref{fig:EPICima}). Its derived X-ray position is also shifted by about the same amount and direction with respect to the Gaia DR3 position of its identified counterpart (see Table\,\ref{tab:xmmastrometry}). Due to the small angular distance between the two sources field distortions and rotation can be neglected. Therefore, we used \nearsxp to compute a linear astrometric correction of -0.44\arcsec\ and +0.52\arcsec\ in R.A. and Dec. for the X-ray position of \src. The resulting final X-ray coordinates for \src are presented in Table\,\ref{tab:xmmastrometry}, fully consistent with the Gaia position within 0.1\arcsec.

%--------------------------
%--------------------------

For the extraction of EPIC spectra and light curves we defined circular regions with radii of 30\arcsec\ and 60\arcsec\ for the source and a nearby source-free background region, respectively. The EPIC spectra were grouped to a minimum
of 20 counts per bin. For the timing analysis of the EPIC-pn data we used the first 11\,ks of the observation, which includes some weak flares but excludes the very strong background flares during the last part of the observation. The source shows little flux variations during the observation.

%%%%% XMM-Newton spectra
\subsubsection{EPIC spectral analysis}
\label{sec:EPICspec}

\begin{figure}
\centering
 \resizebox{\hsize}{!}{\includegraphics{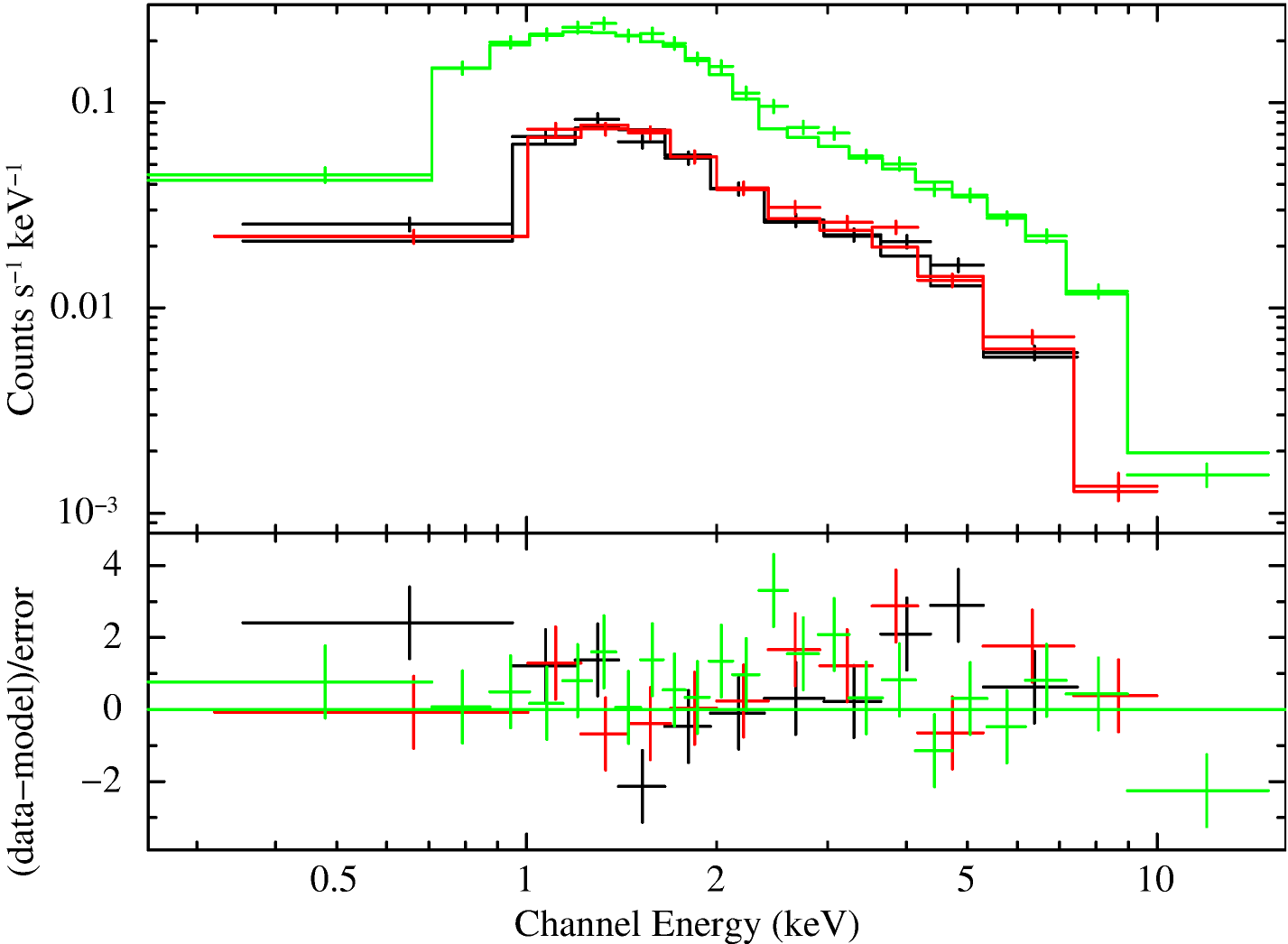}}
  \caption{Top: \xmm EPIC spectra of \src with best-fit model as histograms (MOS1: black, MOS2: red, pn: green). Bottom: Residuals in error units.}
  \label{fig:xspec}
\end{figure}

The X-ray spectrum of \src during the \xmm observation was analysed by simultaneously fitting the model from above to the three EPIC spectra. 
Again, a cross-normalization constant was included which was fixed at 1.0 for the MOS1 spectrum and allowed to vary freely for the other two spectra.
This model describes the spectra well with a $\chi^2$ of 469.0 for 440 degrees of freedom. 
The EPIC spectra with their best-fit model are presented in Figure\,\ref{fig:xspec} and the best-fit model parameters are summarised in Table\,\ref{tab:specfit}.
The observed flux (0.2$-$10.0\,keV, averaged over the three instruments) was 3.43\ergcm{-12} which can be converted to an absorption corrected luminosity of 1.85\ergs{36}, assuming a distance of 62\,kpc to the SMC \citep{2020ApJ...904...13G}.

%--------------------------
%--------------------------

%--------------------------
\begin{table*}
\centering
\caption[]{\xmm and \swift spectral analysis results of \src. Errors on spectral fit parameters are given for 90\% confidence intervals.}
\label{tab:specfit}
\begin{tabular}{lccccccc}
\hline\hline\noalign{\smallskip}
\multicolumn{1}{l}{Instrument} &
\multicolumn{1}{c}{\nhGal} &
\multicolumn{1}{c}{\nhsmc} &
\multicolumn{1}{c}{Photon} &
\multicolumn{1}{c}{F$_{\rm x}$$^{a}$} &
\multicolumn{1}{c}{L$_{\rm x}$$^{a}$} &
\multicolumn{1}{c}{$\chi^2_r$/dof}\\
\multicolumn{1}{l}{} &
\multicolumn{1}{c}{(\ohcm{21})} &
\multicolumn{1}{c}{(\ohcm{21})} &
\multicolumn{1}{c}{index} &
\multicolumn{1}{c}{(erg cm$^{-2}$ s$^{-1}$)} &
\multicolumn{1}{c}{(erg s$^{-1}$)} &
\multicolumn{1}{c}{} \\
\noalign{\smallskip}\hline\noalign{\smallskip}
\xmm EPIC  &  0.56 fixed & 6.4 $\pm$ 0.6       & 1.25 $\pm$ 0.04 & 3.43\expo{-12} & 1.85\expo{36} & 1.07/440\\
\swift-XRT &  0.56 fixed & 5.0$^{+6.0}_{-3.5}$ & 0.81 $\pm$ 0.27 & 1.43\expo{-11} & 7.03\expo{36} & 0.54/14\\
\noalign{\smallskip}\hline
\end{tabular}

\smallskip$^{a}$Observed X-ray flux and absorption-corrected  luminosity in the 0.2$-$10\,keV band assuming a distance of 62\,kpc.
\end{table*}
%--------------------------

\subsubsection{EPIC-pn timing analysis}
\label{sec:pulsations}

\begin{figure}
\centering
 \resizebox{0.95\hsize}{!}{\includegraphics{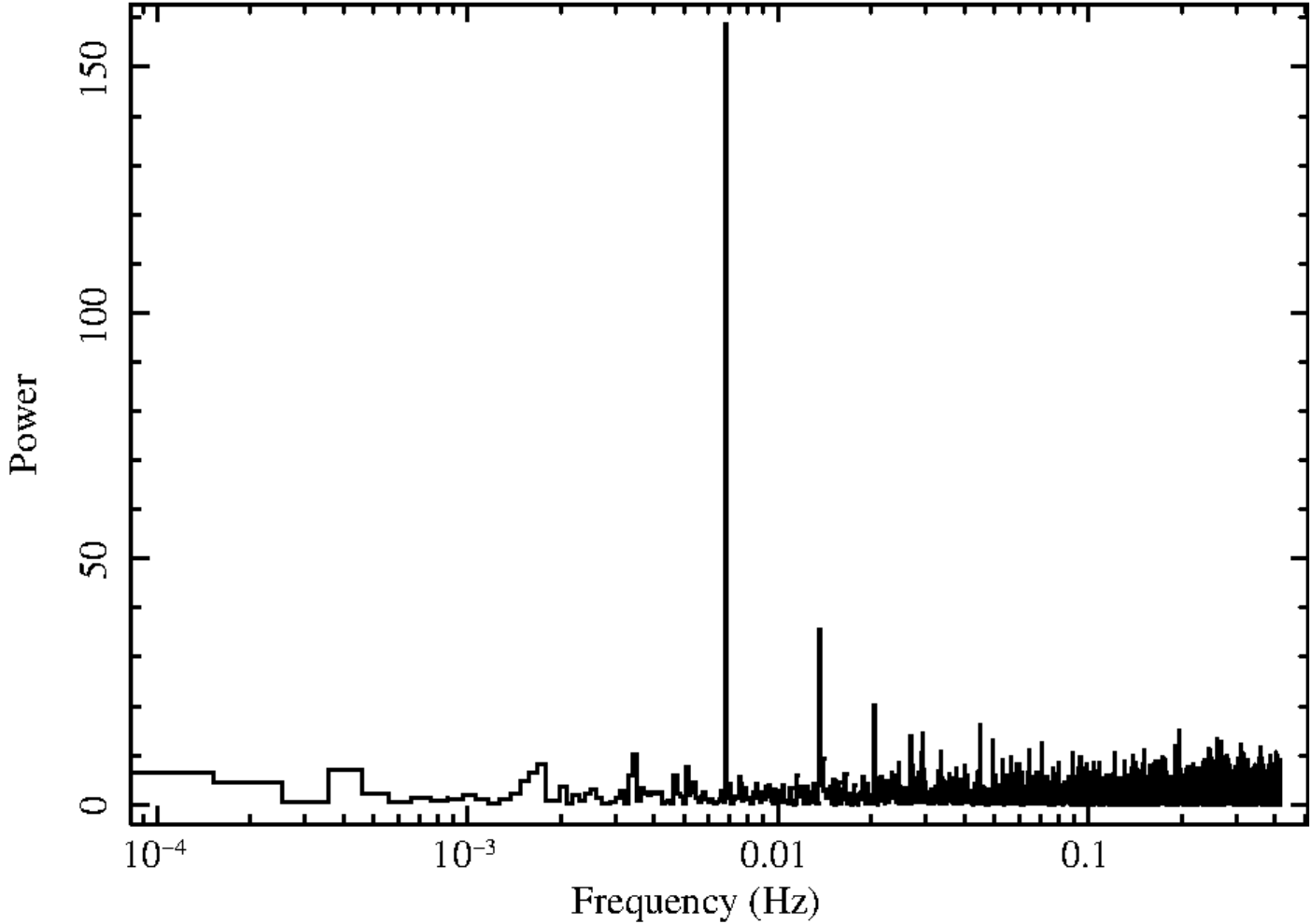}}
  \caption{Power spectrum of \src derived from the 0.2$-$8.0\,keV EPIC-pn light curve with 1.2\,s binning.}
  \label{fig:PNpower}
\end{figure}

To investigate short-term variability of \src, we created EPIC-pn light curves in the 0.2$-$8.0\,keV energy band from the source extraction region. We selected valid pixel patterns (\pattern 1--12) and corrected the event arrival times to the solar system barycentre.
For our search for X-ray pulsations in the light curves, we first created power spectra which revealed pulsations at a frequency of about 6.8\,mHz. 
The power spectrum obtained from the 0.2$-$8.0\,keV EPIC-pn light curve is presented in Figure\,\ref{fig:PNpower}. Besides the fundamental frequency, also the first harmonic at about 13.6\,mHz is visible.

Following recent work on other HMXB pulsars in the Magellanic Clouds \citep[e.g.][]{2022A&A...662A..22H,2017MNRAS.470.4354V}, we determined the precise period and its error in a second step. For this we used the Bayesian approach described by \citet{1996ApJ...473.1059G} in a restricted frequency range around the peak detected in the power spectrum. This results in a pulse period of 146.79 $\pm$ 0.03\,s.

\begin{figure}
\centering
 %\resizebox{0.95\hsize}{!}{\includegraphics{figures/EPICpn_efold.pdf}}
 \resizebox{0.95\hsize}{!}{\includegraphics{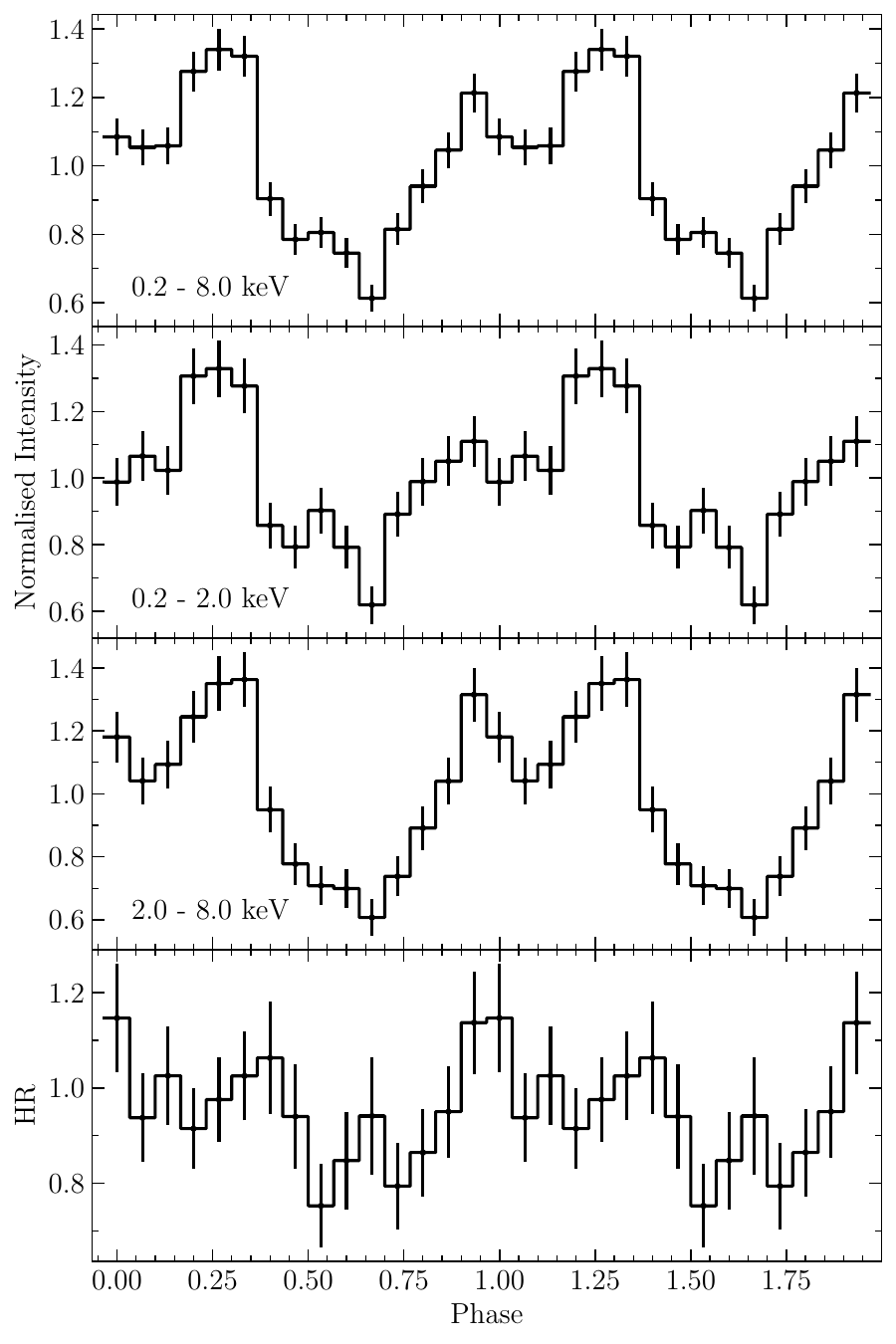}}
  \caption{Pulse profiles, normalised to the average count rates, obtained from folding the EPIC-pn light curves in broad, soft and hard energy bands (top three panels). The hardness ratio (HR, bottom) was derived as count ratio of hard to soft bands.}
  \label{fig:PNefold}
\end{figure}

\begin{figure}
\centering
 \resizebox{1.0\hsize}{!}{\includegraphics{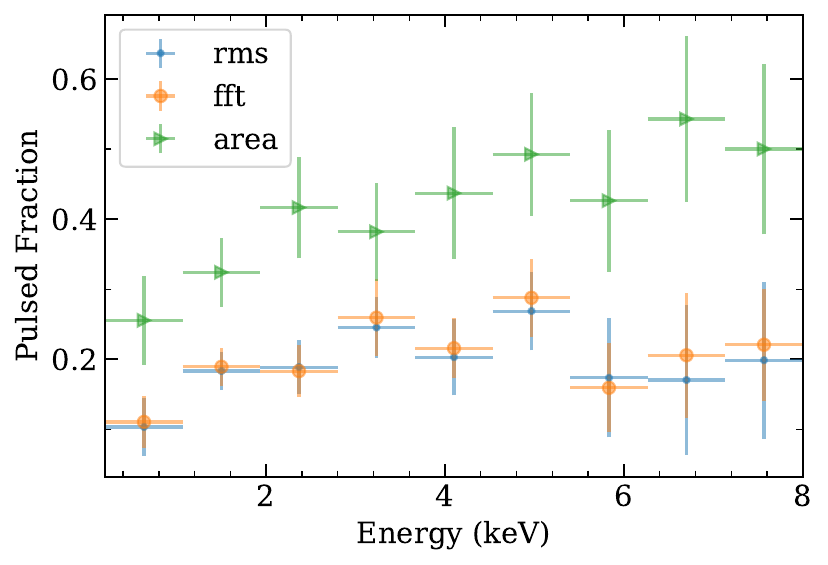}}
  \caption{Pulse profile as a function of energy. The pulsed fraction is computed using three methods outlined in \citet{2023A&A...677A.103F}.}
  \label{fig:PF}
\end{figure}

The pulse profiles in different energy bands, together with the corresponding hardness ratio (HR), are shown in Figure\,\ref{fig:PNefold}. As already indicated by the presence of the first harmonic in the power spectrum, the pulse profile is double-peaked. 
The earlier of the two peaks is more pronounced at energies above 2\,keV, which leads to a maximum in the HR, while lower HRs are indicated during the intensity minimum. 
We infer a pulsed fraction of 36 $\pm$ 8\% (0.5$\times$(maximum - minimum)/mean) from the broad (0.2$-$8.0\,keV) band. 
We computed the pulsed fraction as a function of energy following \citep{2023A&A...677A.103F}. In particular, we used the FFT, RMS, and area methods and computed the pulsed fraction for nine energy ranges between 0.2 and 8.0 keV, with the results shown in Figure\,\ref{fig:PF}. The results reveal somewhat smaller pulsed fraction values below 1 keV, but no significant features are identified, in contrast to other bright outbursts in the Magellanic Clouds \citep{2025MNRAS.536.1357Y}.

%--------------------------
%--------------------------
%--------------------------
%--------------------------

%--------------------------
%--------------------------

\subsection{Long-term X-ray behaviour}
\label{sec:ltflux}

We used the HIgh-energy LIght curve GeneraTor \citep[HILIGT;][]{2022A&C....3800529K,2022A&C....3800531S}\footnote{\url{http://xmmuls.esac.esa.int/upperlimitserver/}} to investigate the long-term X-ray light curve of \src.
The tool uses available catalogues to search for source detections and provides upper limits (2$\sigma$) when the source was not detected. We excluded \ROSAT data because of the limited energy band and spatial resolution. 
We mainly obtained upper limits from archival observations performed with \swift and \xmm before September 2024. To determine the conversion factor for \swift-XRT count rates (0.2$-$12\,keV) to fluxes (0.2$-$10\,keV) we downloaded the spectrum obtained on 2024-09-05 16:12-18:08 UT (ObsID 00016793001, exposure 2.50\,ks), when the flux measured by \swift was highest, from the \swift archive\footnote{\url{https://www.swift.ac.uk/user_objects/index.php}} \citep{2009MNRAS.397.1177E} and modelled it consistently to the EPIC spectra. The best-fit model parameters are listed in Table\,\ref{tab:specfit}. This resulted in a conversion factor of 1.02\ergcm{-10} / (cts s$^{-1}$). For \xmm we used the fluxes provided by HIGLIGT (assuming a power law with photon index 1.5 and an absorption column density of \ohcm{21}). To convert the HILIGT 0.2$-$12\,keV fluxes to 0.2$-$10.0\,keV band we simulated an EPIC-pn spectrum with the HILIGT model parameters and normalised it to the count rate obtained from the observed spectrum.  

\begin{figure*}
\centering
 \resizebox{0.95\hsize}{!}{\includegraphics{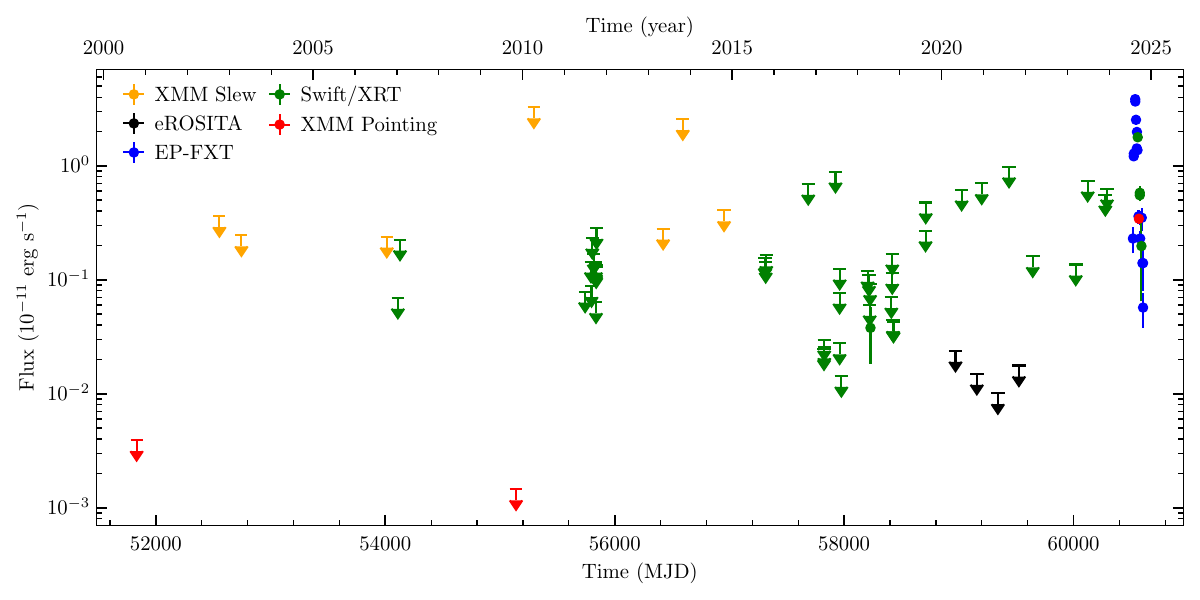}}
 \caption{Long-term X-ray flux (observed, 0.2$-$10.0\,keV) evolution of \src. Down-arrows mark 2\,$\sigma$ upper limits. The evolution of the X-ray outburst with expanded time axis is shown in Figure\,\ref{fig:epflux}.}
  \label{fig:ltflux}
\end{figure*}

Because \src was not detected by \ero \citep{2021A&A...647A...1P}, we determined upper limits from the four \ero surveys, which covered the source. Due to the vicinity of \nearsxp and the size of the point spread function of $\sim$30\arcsec\ \citep[half energy width in survey mode,][]{2024A&A...682A..34M} \nearsxp contributes to the flux at the position of \src.  Therefore, we extracted source and background spectra, with the background region at the same distance to \nearsxp, but on the opposite side of \src in order to take the contribution from this brighter source into account. Assuming the best-fit spectral model from the EPIC spectra, normalised to the 2$\sigma$ count rate upper limits inferred from the spectra, we computed fluxes in the 0.2$-$10.0\,keV band.
Finally, the 0.2$-$10.0\,keV fluxes measured by \EP and from the new \xmm observation were included in the light curve, which is plotted in Figure\,\ref{fig:ltflux} (see also Figure\,\ref{fig:epflux}). It reveals the first strong X-ray outburst seen from \src with a maximum flux a factor of at least 2340 higher than the most stringent upper limit inferred from an \xmm observation in 2009. 

%--------------------------
%--------------------------

\section{Discussion}
\label{sec:discussion}

After the discovery of a new X-ray transient in the SMC by \EP, we  observed \src with \xmm during its outburst. The \xmm observation was performed 31 days after the maximum source flux was observed by \EP, when the flux had declined by a factor of $\sim$10.
The EPIC data reveal a highly significant coherent signal at a period of 146.79 $\pm$ 0.03\,s, which is close to the maximum of the second peak in the pulse-period distribution of BeXRB pulsars in the SMC \citep{2016A&A...586A..81H}. 
The light curve folded at the pulse period shows a double-peaked profile with pulsed fraction of 36\,\%, also typical for BeXRBs. 

Based on the \EPFXT monitoring light curve (see Figure\,\ref{fig:epflux}), the outburst of \src followed a fast-rise-slow-decay profile, lasting more than 90 days. From the \EPFXT spectral analysis results, the absorption corrected X-ray flux in the 0.2--10 keV band reached at least, $L_{\rm peak}$$\sim$1.8\ergs{37}, near the peak of the outburst ($L_{\rm peak}$/$L_{\rm Edd}$$\sim$0.1, here the Eddington luminosity for a neutron star is taken as 1.7\ergs{38}), indicating that \src likely went into a type-II outburst of a transient BeXRB despite the peak luminosity being at the lower end of the distributions of typical values \citep{2011Ap&SS.332....1R}. \EPFXT observed \src in outburst covering the change over nearly two orders of magnitude in luminosity, enabling us to investigate the relation of its spectral properties with luminosity.

\begin{figure}
\centering
 \resizebox{\hsize}{!}{\includegraphics{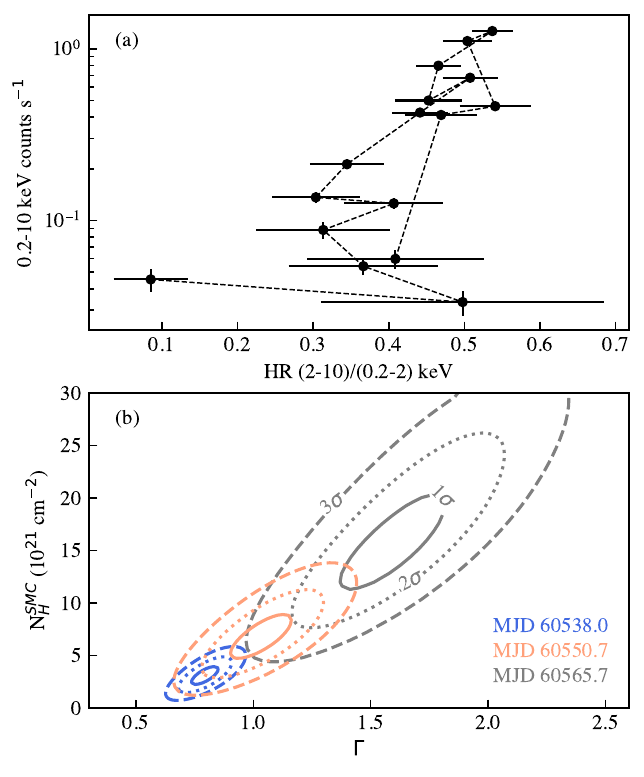}}
  \caption{The hardness-intensity diagram measured by \EP during the outburst of EP J005146.9-730930 (a). The hardness ratio is defined as the ratio of the count rates in the energy band of 2--10 keV and 0.2--2 keV. The quoted numbers are the average measurements by FXTA and FXTB. Confidence contours of the absorption column density, $N_{\rm H}^{\rm SMC}$, and the photon index, $\Gamma$ determined from three representative \EP observations on MJD 60538.0, 60550.7, and 60565.7 (b).}
  \label{fig:ep_hr}
\end{figure}

The spectral evolution of BeXRBs during type-II outbursts has been more thoroughly studied for Galactic sources \cite[e.g.,][]{reig08, tam22, thal24}. The spectral behaviour of \src is consistent with the typical cases known for Galactic sources accreting at relatively low Eddington ratios, in the sense that the energy spectra hardened when the source luminosity became higher (see Figure\,\ref{fig:epspec}e). Considering the most tightly constrained data points, we find that the photon index measured by \EPFXT, $\Gamma$$\sim$0.7--0.8, around the peak of the outburst is significantly smaller than that obtained based on \xmm, $\Gamma$=$1.25\pm0.04$. The decrease of photon index and the increase of absorption column density could both cause spectral hardening. The hardness-intensity diagram of EP J005146.9-730930 during outburst measured by \EP in Figure \ref{fig:ep_hr}a displays a similar trend as Figure \ref{fig:epspec}e, indicating that the spectral hardening behaviour towards high luminosities is model-independent. Also, we plot the confidence contours in Figure \ref{fig:ep_hr}b to investigate the level of parameter degeneracy between $N_{\rm H}^{\rm SMC}$, and $\Gamma$. As an example, the best fit of the \EP data on MJD 60538.0 prefers both lower $N_{\rm H}^{\rm SMC}$ and smaller $\Gamma$ when compared to that on MJD 60565.7 above $3\sigma$ confidence, with a difference of about ten times in luminosity.

The spectral evolution of \src supports the general picture of an accreting neutron star below a certain critical luminosity that is largely determined by the pulsar magnetic field, when the deceleration of the accreting flow at the vicinity of the neutron star surface is dominated by Coulomb interactions rather than radiation pressure \citep[][]{basko76}. Under such circumstances, the increase in luminosity is believed to lead to a decrease in the height of the emission zone in the accretion column above the neutron star surface, hence larger optical depth for the Comptonisation process and producing harder photons \citep{becker12,reig13}. For an approximate estimate, assuming the critical luminosity, $L_{\rm crit}$>$L_{\rm peak}$, and a relation of the critical luminosity with the neutron star surface magnetic field\footnote{The critical luminosity is derived to be 1.5$\times$10$^{37}B_{12}^{16/15}$ erg s$^{-1}$ in \citet{becker12}. And here we multiply the absorption-corrected luminosity in the 0.2--10 keV band by a factor of 3 to estimate the bolometric luminosity of \src.}, we could obtain a lower limit for the magnetic field of \src as, $B$>$3.3\times10^{12}$G.

There are also hints for a tentative anti-correlation between the absorption column density and the source luminosity (see Figure\,\ref{fig:epspec}f), which could be caused by absorption variations related to orbital geometrical effects, similar as observed during a recent giant outburst of the Galactic BeXRB EXO 2030+375 \citep{thal24}. We note that the measurements of N$_{\rm H}^{\rm SMC}$ derived from \src are subject to large uncertainties, as only upper limits are yielded by the last few \EP observations. Based on only the data points with constrained measurements from \EP, \xmm, and \swift, the standard Pearson correlation coefficient between the absorption column density, $N_{\rm H}^{\rm SMC}$, and the logarithmic X-ray luminosity, $L_{\rm x}$, is found to be -0.746, with a \textit{p}-value of 0.005, indicating an anti-correlation with a confidence of $\sim$$2.79\sigma$.

%--------------------------
%--------------------------

During many serendipituous X-ray observations before 2024 \src was never significantly detected. Many short observations yielded relatively non-constraining 2\,$\sigma$ upper limits. Formally, only one measurement was indicated as detection, but the flux is again only twice its error. Two \xmm pointed observations provide stringent upper limits about three orders of magnitude below the maximum observed during outburst. Duration, high amplitude and maximum X-ray luminosity of $\sim$2\ergs{37} all suggest that \src was for the first time detected in X-rays during a type-II outburst.

\section*{Acknowledgements}

This work is based on data obtained with \fullEP and \xmm. 
\fullEP is a space mission supported by the Strategic Priority Program on Space Science of Chinese Academy of Sciences, in collaboration with ESA, MPE and CNES (grant no. XDA15310000), the Strategic Priority
Research Program of the Chinese Academy of Sciences (grant no. XDB0550200), and the National Key R\&D Program of China (2022YFF0711500). Y.X. acknowledges  support by the Hundred Talents Program of the Chinese Academy of Sciences.  C.J. acknowledges the National Natural Science Foundation of China through grant 12473016. LD acknowledges funding from German Research Foundation (DFG), Projektnummer 549824807. \xmm is an ESA science mission with instruments and contributions directly funded by ESA Member States and NASA. The \xmm project is supported by the DLR and the Max Planck Society. This work made use of data supplied by the UK Swift Science Data Centre at the University of Leicester.

%%%%%%%%%%%%%%%%%%%%%%%%%%%%%%%%%%%%%%%%%%%%%%%%%%
\section*{Data Availability}

The data presented in the tables and figures of the paper are available upon reasonable request from the corresponding author. 
The data from the \xmm observation will become available from the \xmm Science Archive\footnote{\url{https://www.cosmos.esa.int/web/xmm-newton/xsa}} at the end of September 2025.



\bibliographystyle{mnras}
\bibliography{general}

% Don't change these lines
\bsp	% typesetting comment
\label{lastpage}
\end{document}